\documentclass[aps,prb,preprint,superscriptaddress,showpacs,tightenlines]{revtex4-1}
\usepackage{epsfig,amssymb,amsfonts,amsmath}
\begin{document}

\preprint{Pre-print of \emph{Nature Physics} (2017) DOI:10.1038/NPHYS4175 }

\title{Long-distance spin transport in a disordered magnetic insulator}


\author{Devin  Wesenberg}
\affiliation{Department of Physics and Astronomy, University of Denver, Denver, CO 80208}
\author{Tao Liu} 
\affiliation{Department of Physics, Colorado State University, Fort Collins, CO  80523}
\author{Davor Balzar}
\affiliation{Department of Physics and Astronomy, University of Denver, Denver, CO 80208}
\author{Mingzhong Wu}
\affiliation{Department of Physics, Colorado State University, Fort Collins, CO  80523}
\author{Barry L. Zink}
\affiliation{Department of Physics and Astronomy, University of Denver, Denver, CO 80208}

\date{\today}

\begin{abstract}
Spin transport through magnetic insulators via magnons has recently been explored for a growing variety of magnetic systems with long-range order and well-understood spin excitation spectra.  Here we show dramatic effects of spin transport through an amorphous magnetic insulator, which is both magnetically and structurally disordered.  We generate and detect spin flow though amorphous yttrium-iron-garnet ($a$-YIG) thin films in a non-local geometry by use of the spin Hall and inverse spin Hall effects in platinum strips separated by 10 or more microns.   By comparing non-local spin transport in  $a$-YIG on suspended micromachined thermal isolation platforms to the same experiment performed on a bulk substrate, we show strong effects of in-plane thermal gradients on spin transport in the disordered magnetic insulator.  The resulting non-local voltage signals are orders of magnitude larger than those seen in crystalline magnetic insulators, with easily measurable spin signals seen even at distances in excess of $100$ microns.  In analogy to heat transport, where disordered materials support a range of vibrational excitations that can allow large thermal conductivities, we suggest that efficient spin transport in disordered magnetic systems can occur via a similar spectrum of excitations that relies on strong local exchange interactions and does not require long-range order.  This work not only opens a new area for fundamental experimental and theoretical studies of spin transport, but also sets a new direction in materials science for magnonic and spintronic devices. 

\end{abstract}


\maketitle

Motivated by new paradigms for information processing, spintronics research has recently focused on the transport of spin information via spin-wave, or magnon, excitations in magnetic insulators.  Much of this work uses yttrium iron garnet, Y$_{3}$Fe$_{5}$O$_{12}$ (YIG), as the spin transport medium due mostly to its very low damping of magnetization dynamics and the resulting long spin-wave propagation lifetime.\cite{ChumakNatPhys2015,SergaJPhysD2010}  In its bulk crystalline form YIG is a ferrimagnet with an electronic bandgap of $\approx2.8\ \mathrm{eV}$, which is also achieved in thin films,\cite{JakubisovaJAP2015} so that electronic excitations can not contribute to transport.  The ferrimagnetism arises due to the location of Fe$^{3+}$ ions in two inequivalent sites in the relatively complicated unit cell, leading to antiferromagnetic exchange interactions between octahedrally- and tetrahedrally-coordinated Fe$^{3+}$ ions but with somewhat different moments, leaving a net imbalance of magnetization and macroscopic properties often described using the typical language of ferromagnets.  

\begin{figure*}
\includegraphics[width=6.0in]{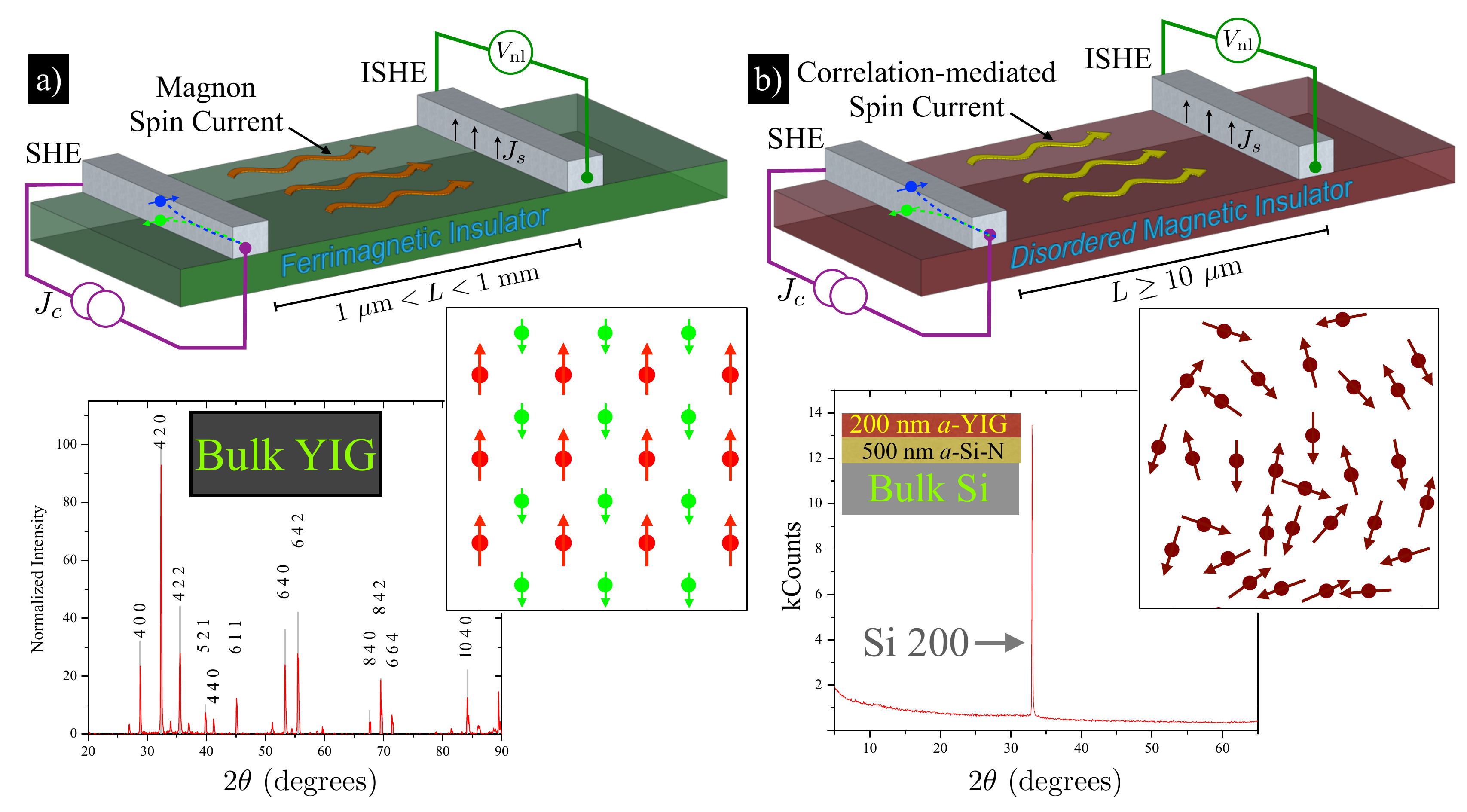}
\caption{Schematic views of experiments in long-distance spin transport. \textbf{a)} Non-local spin transport in crystalline YIG, a ferrimagnetic insulator.  Charge current driven through a platinum strip causes a spin current in the Pt thickness direction flow via the spin Hall effect (SHE), generating spin torque and/or spin accumulation at the Pt/YIG interface, exciting magnons that carry spin information to a Pt detector where spin current injected generates a charge voltage via the inverse spin Hall effect (ISHE).  The X-ray diffraction (XRD) pattern shows Bragg reflections consistent with randomly oriented polycrystalline YIG.  A simplified 2D schematic spin structure of YIG shows two sublattices of Fe spins, where nearest neighbor interactions are antiferromagnetic.  \textbf{b)} Spin transport through a disordered magnetic insulator, $a$-YIG, relying only on magnetic correlations.   XRD on a 200 nm thick YIG layer sputtered on a Si substrate coated with 500 nm of $a$-Si-N shows no YIG diffraction peaks, indicating the lack of any medium- or long-range order in the YIG layer.  Strong peak at $\sim35^{\circ}$ is due to the Si substrate ($200$) Bragg reflection.  A simplified 2D schematic random spin structure of $a$-YIG, illustrates the high degree of frustration, and lack of long-range order despite strong AF interactions between neighboring spins. }
\label{Cartoon}
\end{figure*}

Among the most exciting spin transport studies in YIG are experiments demonstrating electrical excitation of spin waves in the YIG via the spin Hall effect (SHE),\cite{SinovaRMP2015,HoffmannIEEETransMag2013,HirschPRL99,DyakonovPLA1971} and subsequent detection of the spin information some distance away from the injection site via the reciprocal inverse spin Hall effect (ISHE).  This non-local generation and detection of spin information transport in YIG, shown schematically in Fig.\ \ref{Cartoon}a, was first described by Kajiwara, \emph{et al}. \cite{KajiwaraNature2010} where very long length scale propagation was claimed. Only recently have other reports of similar experiments also on YIG emerged, showing shorter propagation length scales.\cite{CornelissenPRB2016,VelezPRB2016,CornelissenNatPhys2015,GoennenweinAPL2015,GilesPRB2015}  All of these experiments focus on crystalline or epitaxial YIG, though depending on the process steps used in fabrication some level of disorder could arise.  The study by Kajiwara, \emph{et al}. also reported the excitation of magnetization dynamics in crystalline YIG by SHE-driven torques, a phenomenon also recently reported using different device structures,\cite{LauerAPL2016,JungfleischPRL2016,SafranskiArxiv20016,HamadehPRL2014,ChumakAPL2012} including some that deliberately enhance the role of thermal gradients.\cite{SafranskiArxiv20016}  The SHE excitation of YIG magnetization has also been theoretically described.\cite{ZhouPRB2013,XiaoPRL2012,ChenPRL2015}  The characteristic feature of this SHE spin-wave excitation is an onset of the dynamics at a critical current where the applied spin torque balances the damping of the magnetization dynamics in the YIG.  
Furthermore, thermal gradients and spin-wave excitations have been shown to have other dramatic interactions in YIG,\cite{CunhaPRB2013,AnNatMater2013} opening the possibility that the application of thermal gradients in these experiments, whether intentional or unintentional, could play a strong role in measured effects.  This has been observed by some groups,\cite{CornelissenPRB2016,CornelissenNatPhys2015} though the applied thermal gradients tested to date are overwhelmingly perpendicular to the plane of the YIG/Pt interfaces.  Nevertheless, the thermal generation of a population of magnons that subsequently diffuses through the YIG is one possible mechanism for the long-distance spin flow.\cite{ShanPRB2016}  

Other recent reports have shown that spin transport is possible through a much wider range of materials than previously thought.  These include studies of spin transport, and possible enhancement of spin flow, through very thin nickel oxide\cite{WangPRL2014AF,HahnEPL2014,LinPRL2016} and other nominally antiferromagnetic insulating\cite{WangPRB2015} layers inserted between YIG and Pt layers, and through thin native oxides of nickel and Permalloy between transition metal ferromagnets and heavy metal films.\cite{ZinkPRB2016}  These initially unexpected experimental results have stimulated theoretical consideration of spin transport by magnons in antiferromagnetic insulators.\cite{KhymynPRB2016,BaltzarXiv2016,RezendePRB2016} In addition to these studies, where spin transport was shown via electrically-detected measurements of the ISHE in response to spin pumping, the longitudinal spin Seebeck effect has been demonstrated in antiferromagnets, \cite{WuPRL2016,PrakashPRB2016,SekiPRL2015} paramagnets,\cite{WuPRL2015} and ferromagnets above the Curie temperature.\cite{ShiomiPRL2014}   These results clearly demonstrate that long-range magnetic order is \emph{not} a requirement for spin transport in an insulator, which is also implicit in any spin transport experiment using a very thin film of a material that is antiferromagnetic \emph{in bulk}, but with a blocking temperature well below the temperature of the experiments.\cite{WangPRL2014AF,HahnEPL2014,WangPRB2015,ZinkPRB2016}
New experiments to test a broader range of disordered magnetic insulators, where magnetic correlations persist due to strong local exchange interactions despite the lack of a low symmetry state, are therefore critical for spintronics.

In this paper we show that a disordered magnetic insulator allows long-distance spin transport.          
We demonstrate non-local spin transport (see Fig.\ \ref{Cartoon}b), with large signal voltages indicating propagation over dozens of microns, through \emph{amorphous} YIG ($a$-YIG), a magnetic insulator with strong local antiferromagnetic exchange interactions but neither magnetic nor structural long-range order.  We describe non-local spin transport in $a$-YIG films sputtered both on suspended amorphous Si-N sample platforms and on bulk Si substrates.  Comparing these allows us to identify a strong effect on in-plane thermal gradients.  We show two separate contributions to the non-local spin transport, with one showing a clear onset at well-defined critical current density in the Pt across a fairly broad range of samples and measurement conditions, while the other is linear with applied current through the strip.    
Finally, when the non-equilibrium spin carriers are injected into $a$-YIG the temperature profile suggests efficient heat transport by this spin population, which echoes the strong magnon-phonon coupling often observed in crystalline YIG.  These results open a new frontier in insulating spintronics, proving that magnetic order is not required, and may not be desirable, for an efficient spin-transport medium.

Amorphous YIG was originally studied, though far from exhaustively, decades ago. Results indicated that disordered YIG (rarely grown in thin-film form) 
showed a broad peak in $M$ vs. $T$ between $50$ and $100$ K,\cite{GyorgyJAP1979,ChukalkinPSSA1989} with a splitting between curves measured in zero field cooled and field cooled conditions.\cite{ChukalkinPSSA1989}  Above this splitting, some groups reported reasonable agreement of $M$ vs. $T$ with a Curie-Weiss law, $M\propto 1/T-\theta$ with a large negative $\theta$ on the order of $100\ \mathrm{K}$ indicating the presence of strong antiferromagnetic (AF) exchange interactions.  Since the expectation for $a$-YIG is that the nearest-neighbor environment is largely unchanged from the crystalline state, local AF interactions are reasonable, though existing reports disagree on this issue. \cite{GyorgyJAP1979,ChukalkinPSSA1989}     The lack of long-range order gives rise to frustration, pushing $a$-YIG toward spin glass or more complex non-equilibrium behavior.  Here one expects strong AF correlations between neighboring spins up to a temperature scale comparable to the bulk transition temperature, with lower temperature freezing phenomena that depend on the balance of the competing interactions in a particular structure.

Using techniques previously shown to produce high-quality epitaxial YIG films when the proper crystalline substrate was used and the proper post-annealing was conducted,\cite{ChangIEEEMagLett2014} we sputtered $100$ nm and $200$ nm thick films of $a$-YIG on amorphous silicon-nitride ($a$-Si-N) coated Si substrates and also on $a$-Si-N thermal isolation platforms\cite{SultanPRB2013} developed for thermal and thermoelectric characterization of thin films and nanostructures.\cite{AveryPRB2015,AveryPRL2012} 
Figure\ \ref{Cartoon} shows x-ray diffraction (XRD) patterns comparing a polycrystalline bulk YIG sample (\textbf{a)}) to an $a$-YIG film on the Si-N coated Si substrate (\textbf{b)}).  The former indicates randomly oriented polycrystalline YIG, whereas the latter exhibits no medium- or long-range order in the YIG layer.  We also performed magnetization measurements of a similar $a$-YIG sample on a Si substrate via SQUID magnetometry.  $M$ vs. $T$ (after subtraction of backgrounds from the substrate and sample mount as described in Supplemental Materials) shows a broad peak near 50 K described in literature\cite{GyorgyJAP1979,ChukalkinPSSA1989} and a second, not previously observed feature near $230$ K discussed further below.


\begin{figure*}
\includegraphics[width=\linewidth]{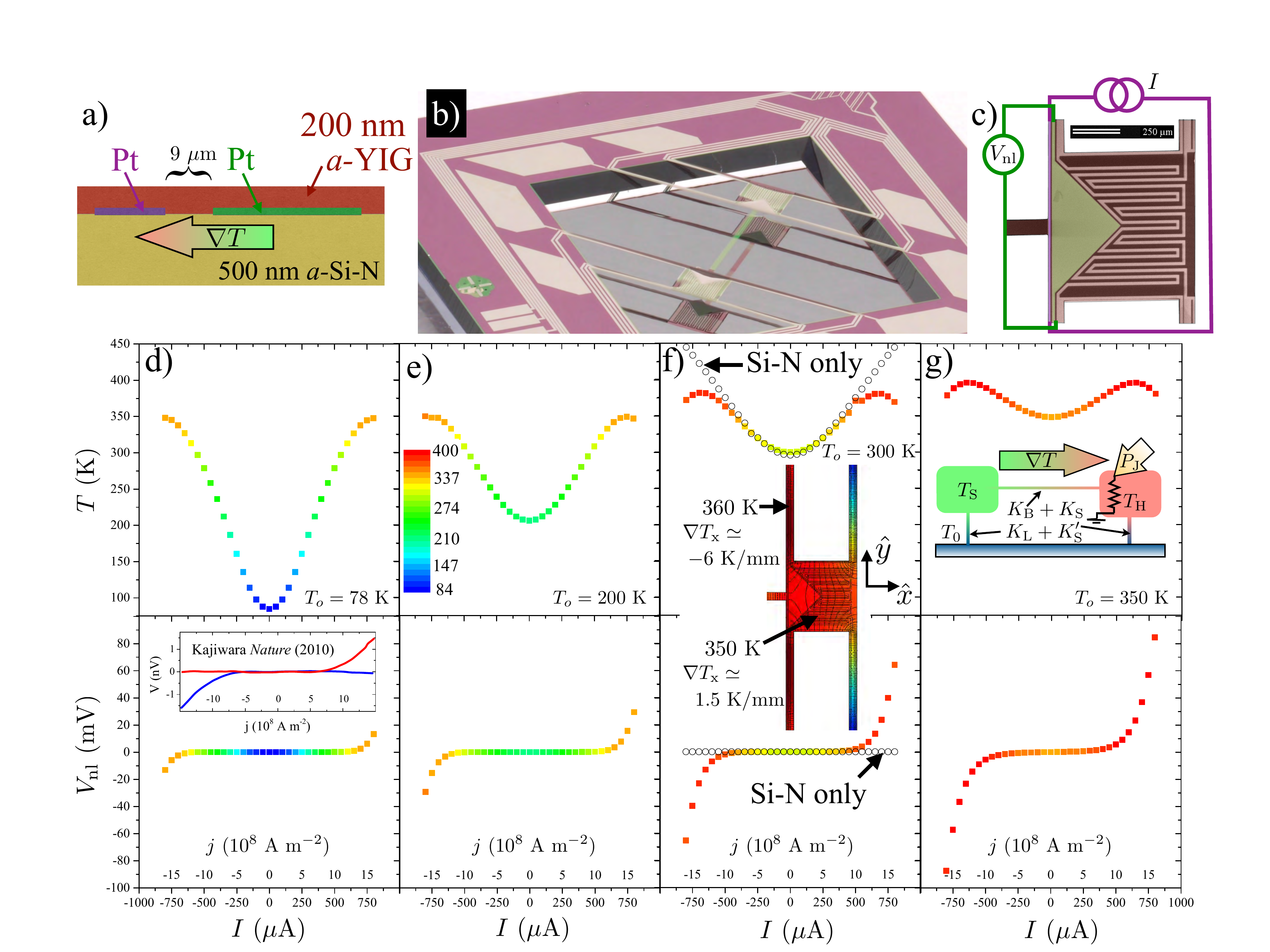}
\caption{Non-local spin transport through suspended $a$-YIG. \textbf{a)} Schematic cross-section of the Si-N platform with $200$ nm of $a$-YIG on the $500$ nm thick Si-N membrane, with locations of injection (purple) and detection (green) Pt strips indicated along with the direction of the thermal gradient. \textbf{b)} Optical micrograph of the thermal isolation platform. \textbf{c)} False color scanning-electron micrograph depicts the suspended non-local spin transport measurement.  Bottom panels \textbf{d-g)} display $V_{\mathrm{nl}}$ vs. $I$ for four different base temperatures (where the colormap indicates the measured $T$ of the island), $T_{o}$.  In each, an abrupt onset of non-local voltage occurs above $500\ \mathrm{\mu A}$ ($10\times 10^{8}\ \mathrm{A/m^{2}}$), with positive $V_{\mathrm{nl}}$ developed for positive $I$ and negative $V_{\mathrm{nl}}$ for negative $I$.  As shown in the inset to \textbf{d)}, a similar pattern was reported for crystalline YIG,\cite{KajiwaraNature2010} though in that case the field must be reversed (red and blue lines) to achieve the opposite polarization of the spin current in the YIG.  In the disordered YIG there is no special direction set by the film magnetization, allowing transport in both channels with no external field.  The open circles in panel \textbf{f)} for $T_{o}=300\ \mathrm{K}$ show the result of the same non-local measurement performed on a Si-N structure with no $a$-YIG layer, and is essentially zero for all $I$, as expected. Each top panel shows the concurrent measurement of the temperature of the Si-N island made via an entirely separate thin film thermometer.  For large $I$ this $T$ first slows its rise with increasing $I$ then for higher $T_{o}$ actually cools due to increased heat transport by spin excitations.  
Again the open circles in panel \textbf{f)} for $T_{o}=300\ \mathrm{K}$ shows $T$ in the absence of $a$-YIG with the simple parabolic dependence expected. We measured the same patterns in a platform coated with $a$-SiO$_{2}$ (see \emph{Supplement}).  Insets to \textbf{f)} show calculated in-plane thermal gradients, as described in the text. Inset to \textbf{g)} shows the simple 2-body model used to determine thermal conductance.}
\label{aYIGdata}
\end{figure*}

Results from the membrane experiments appear in Fig.\ \ref{aYIGdata}.  First note that the open circles in panel \textbf{f)} for $T_{o}=300\ \mathrm{K}$ result from the non-local measurement performed on a Si-N structure with no $a$-YIG layer, and is essentially zero for all $I$, as expected.   Note also that there is a finite but very small amount of charge current leakage through the YIG (resistance from the injector to the detector is always $>100\ \mathrm{k\Omega}$ at room temperatures and much larger at low temperatures) that is always too small to account for the measured non-local voltages (for additional details see Supplemental Materials).  Across all four measured base $T$, a non-linear component to $V_{\mathrm{nl}}$ with a sign change that rules out simple heating effects, is reminiscent of the pattern seen in the original experiments on magnon spin currents in YIG,\cite{KajiwaraNature2010} keeping in mind that the disordered material has no net magnetization and no preferred direction so that either sense of spin current can propagate.  Despite this similarity, which suggests that a non-equilibrium spin population could become self-oscillatory when enough spin-torque is provided by the SHE, the voltages we measure are \emph{many orders of magnitude larger}.  We also reiterate that this large voltage was measured across a distance of nearly $10$ microns.

Another extremely unusual feature of the data is seen most clearly in Fig.\ \ref{aYIGdata}f) and g), where the measured temperature of the Si-N island coated with $a$-YIG actually drops dramatically with increasing $I$.  The thermometer is measured using an AC technique, which is very unlikely to suffer interference from the large DC current applied to the Pt strip.  This also cannot be due to the Peltier effect,\cite{AveryPRL2013} which would be linear with applied $I$, causing heating with one polarity and cooling with the other.  We believe this large drop in the temperature of the island, which we have observed on multiple platforms and with different $a$-YIG thickness, is driven by the addition of a new channel for heat conduction created in the $a$-YIG in response to the SHE injection of the non-equilibrium spin population.  In other words, a non-equilibrium conductance ,$K_{\mathrm{spin}}$, is added to the thermal conductance of the leg (as defined in the thermal model of Fig.\ \ref{aYIGdata}g).  We estimate this $K_{\mathrm{spin}}$ could exceed the thermal conductance of the $a$-YIG by more than 2 orders of magnitude (for further details see supplemental materials). 

As shown in the inset to Fig.\ \ref{aYIGdata}f and in supplementary materials, we use 2D finite-element analysis software to estimate the size of in-plane thermal gradients generated in the suspended thermal platform during non-local spin transport.  The image depicts $T$ calculated for the condition where $\Delta T=50$ K between the Si frame held at $300$ K and the island thermometer.  Since heat is dissipated in the Pt lead that runs along the entire length of the legs of the structure, the peak $T=360$ K is actually on the leg.  In general, the in-plane thermal gradient along the leg has components along both the $\hat{x}$ and $\hat{y}$ directions, with $\nabla T_{\mathrm{x}}$ reaching absolute values near $6\ \mathrm{K/mm}$ in the region between the two Pt leads at the peak $T$ location, and with a maximum value of $18\ \mathrm{K/mm}$ achieved near the connection to the bulk Si frame.  
Peak values of $\nabla T_{\mathrm{y}}\simeq 70\ \mathrm{K/mm}$ along the legs occur in a similar region.  These much larger gradient areas could dominate the additional heat-sinking via spin excitations that drives the overall cooling of the suspended island.  Finally, we note that $\nabla T_{x}$, which we hypothesize plays a role in increasing the non-local voltage signal, actually varies in magnitude and sign across the structure, suggesting that similar devices optimized to produce large and uniform $\nabla T_{x}$ could lead to even more dramatic spin transport effects in suspended $a$-YIG.    


\begin{figure*}
\includegraphics[width=\linewidth]{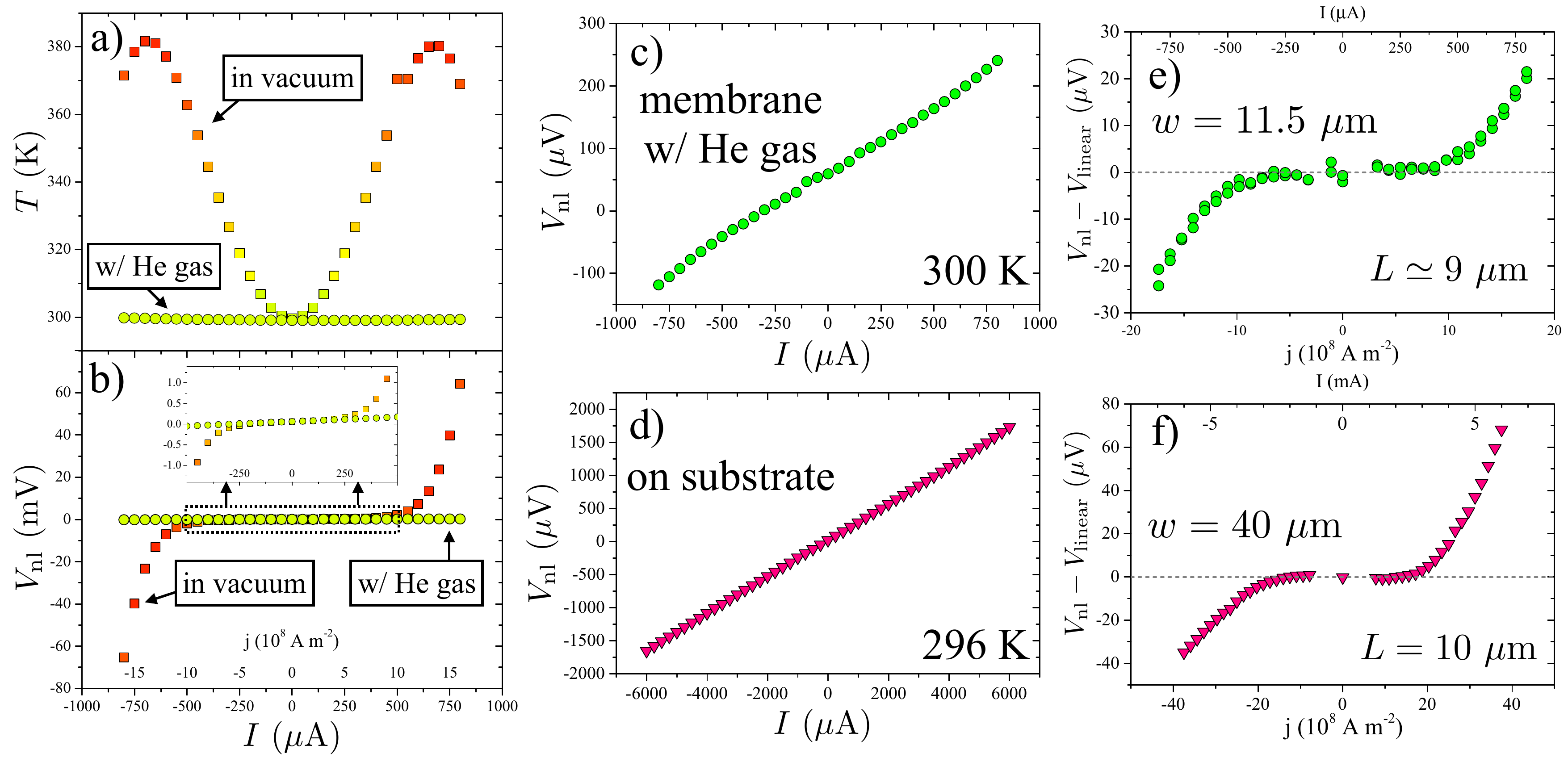}
\caption{Intentional manipulation of direction of thermal gradient.  \textbf{a)} Compares $T$ measured on the island thermometer vs $I$ applied to the Pt spin injector for the $a$-YIG coated Si-N membrane structure with and without He exchange gas surrounding the membrane.  The Helium thermally shorts the structure to the surroundings, such that heating is dramatically reduced, as is the associated lateral thermal gradient between the thermometer, spin detector, and spin injector.  The high thermal conductance of the gas drives any existing thermal gradient \emph{out of the plane} of the sample. As seen in \textbf{b)}, this change dramatically reduces the non-linear component of the non-local signal, though as shown in the inset, a linear component remains essentially unchanged.  \textbf{c)} and \textbf{d)} compare the total non-local voltage measured in the membrane with exchange gas to the $a$-YIG on the bulk Si substrate.  In both cases the thermal gradient created is perpendicular to the film normal as shown in Fig.\ \ref{Fig5}a).  Subtracting the linear term in these plots, as shown in \textbf{e)} and \textbf{f)}, reveals a similar non-linear signal as seen in the membrane, though with reduced signal size.  Note that this non-linear signal turns on at a similar \emph{current density} despite much wider Pt strips used in the substrate experiment.  This suggests the non-linear part is indeed due to spin torque compensated dynamics in the disordered YIG.}
\label{Fig3}
\end{figure*}

Figure \ref{Fig3} describes two different approaches to manipulate the direction of the applied thermal gradient in the non-local spin transport experiment.  First,  we compare $V_{\mathrm{nl}}$ measured in vacuum as in Fig.\ \ref{aYIGdata} with the signal measured on the same sample but with helium gas added to the cryostat to thermally short the Si-N structures to the sample environment.  As is clear from the measured $T$ as a function of $I$ shown in Fig.\ \ref{Fig3}a, in-plane gradients are nearly entirely eliminated, and the dominant gradient is normal to the interface of the heated Pt strip and the gas and therefore very similar to the situation when the Pt/$a$-YIG is supported on a bulk substrate.  As shown in Fig.\ \ref{Fig3}b, this reduces the size of the non-local signal.  Despite the reduced size, both a linear and a non-linear term remain easily measurable when only an out-of-plane gradient exists as shown in Figs.\ \ref{Fig3}c-f.   Figs. \ref{Fig3}c) and d) show the total measured $V_{\mathrm{nl}}$, while e) and f) show the signal after subtraction of the linear term (determined via least-squares fit to the small $I$ region) in order to examine non-linear contributions.  Note that in both experiments, where the exciting Pt strips have very different width, the non-linear $V_{\mathrm{nl}}$ turns on at similar current density, $10\times10^{8}\ \mathrm{A/m}^2 < j < 20\times10^{8}\ \mathrm{A/m}^2$.
This large difference in magnitude of $V_{\mathrm{nl}}$ between in-plane gradient and out-of-plane gradient cases could relate to the presumed large difference in the absolute magnitude of thermal gradients produced in the two experiments.   However these comparisons are complicated by the difficulty in estimating out-of-plane gradients when the constituent materials' thermal properties and nature of the interfaces between them are poorly known.  Here we can use FEM to roughly estimate a value near $0.8\ \mathrm{K/mm}$ for the out-of-plane gradient, with negligible in-plane gradients on distances greater than even one micron away from the Pt current strip.  In our view the most reasonable assumption is that our experiments on the substrate do not involve significant thermal gradients, and instead probe purely electrical spin generation, transport, and detection though further experiments are required to confirm this. 


\begin{figure}
\includegraphics[width=8.6cm]{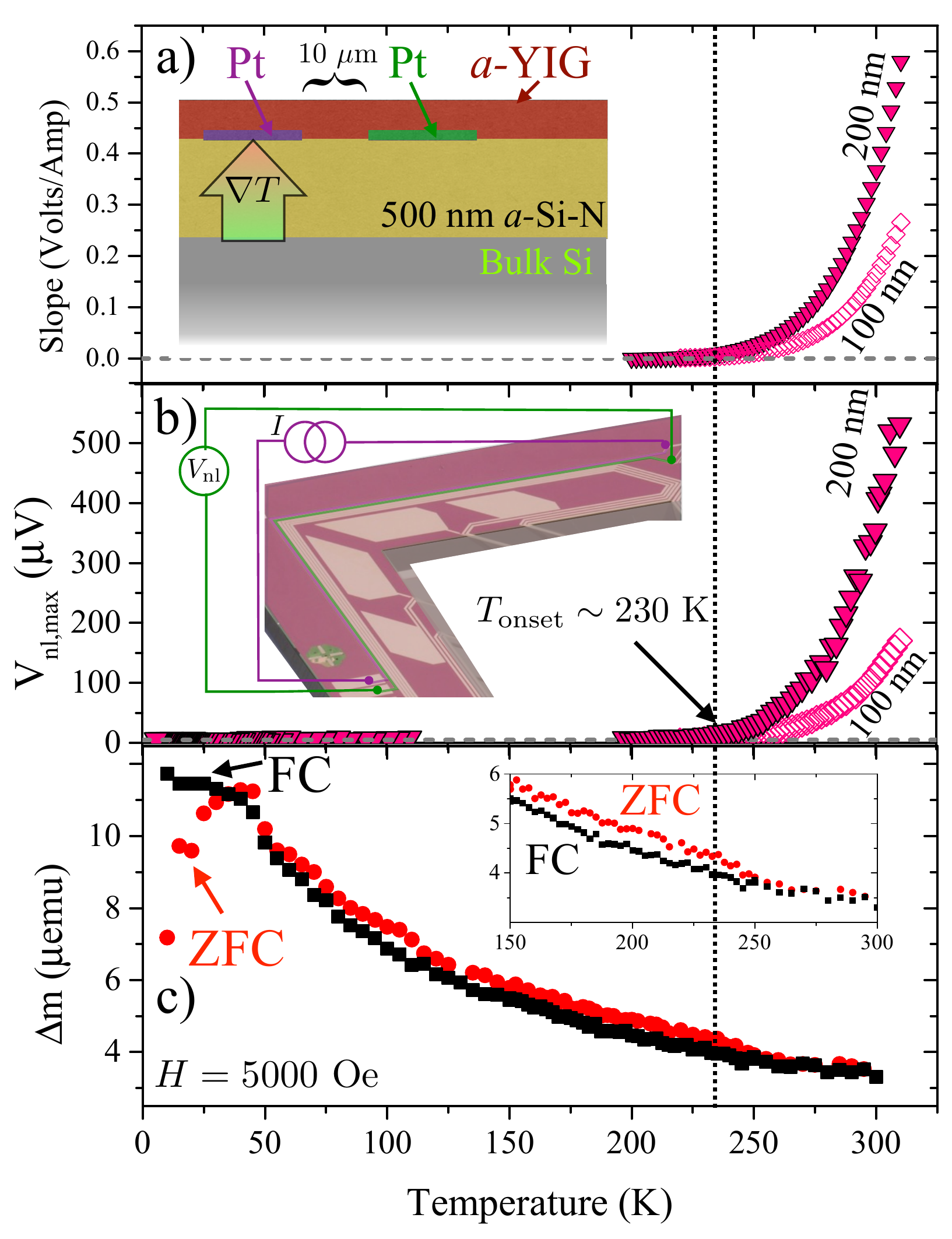}
\caption{ $V_{\mathrm{nl}}$ vs. $T$ from $5$ to $300$ K indicating spin transport through $100$ and $200$ nm thick $a$-YIG on the substrate. \textbf{a)} Linear component (slope) of $V_{nl}$ (driven by purely electrical spin injection and subsequent diffusion).  \emph{Inset}: Schematic view of the non-local experiment.  \textbf{b)} Maximum non-linear $V_{\mathrm{nl}}$ provides an estimate of the component potentially related to ST-driven spin excitations.  Both components of $V_{\mathrm{nl}}$ appear only above $\sim 230\ \mathrm{K}$. \emph{Inset}: Optical micrograph of isolation platform frame showing the location of the substrate-supported non-local measurement.   \textbf{c)} Magnetization of the $a$-YIG vs.\ $T$ from $5$ K to $300$ K cooled in zero field (ZFC, red symbols) and in the 5000 Oe measuring field (FC, black symbols) shows the broad peak near 50 K described in literature\cite{GyorgyJAP1979,ChukalkinPSSA1989}  and a second, not previously discussed feature near 230 K.  This higher-$T$ feature correlates well to the onset of spin transport effects.}
\label{Fig5}
\end{figure}

\begin{figure}
\includegraphics[width=3.38in]{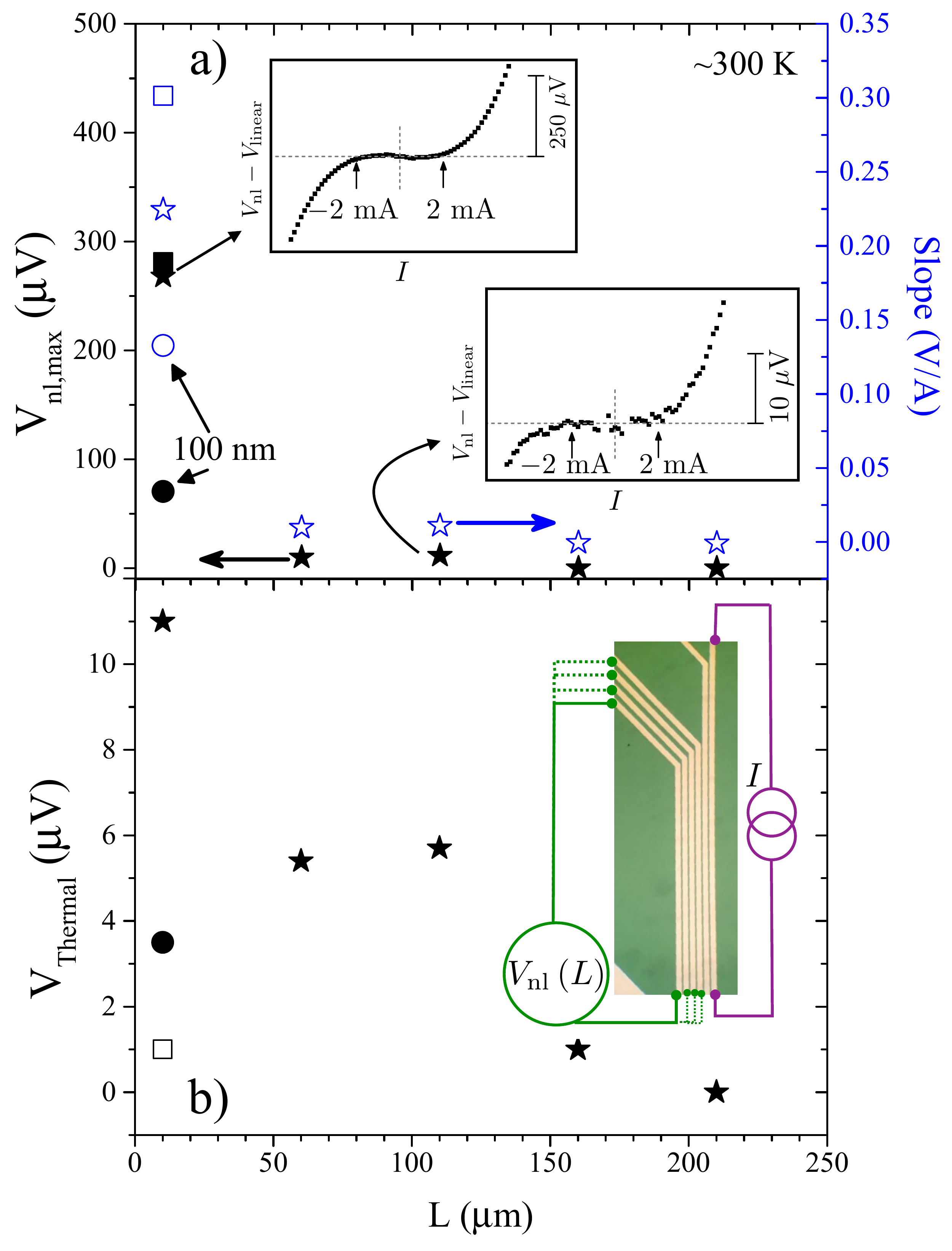}
\caption{ Distance-dependence of $V_{\mathrm{nl}}$ on the substrate. \textbf{a)} Voltage components related to spin drop off sharply with distance.  Here blue symbols indicate slope (right axis) and black symbols maximum non-linear component (left axis).  Stars, boxes, and circles indicate three different samples (\emph{Insets} Nonlinear spin signals after linear subtraction show clear effects even for $L>100\ \mathrm{\mu m}$.) \textbf{b)} Estimation of the (small) heating effects drops off much more slowly, reinforcing that the spin signals are not simply temperature driven but require SHE excitation.}
\label{Fig4}
\end{figure}

Though the largest effects come on the membrane, there the exact temperature of the $a$-YIG transporting spin is difficult to discuss.  In light of this we explore the $T$- and $L$-dependence of the effect in detail using the substrate-supported case, as shown in Figs.\ \ref{Fig5} and \ref{Fig4}.  Fig.\ \ref{Fig5}a) shows the component of $V_{\mathrm{nl}}$ purely linear in $I$ (determined from fits to the slope of $V_{\mathrm{nl}}$ vs $I$ at each $T$) for both $100$ nm and $200$ nm thick $a$-YIG films.  Fig.\ \ref{Fig5}b) shows the maximum recorded value of the non-linear component (here taken at $I=8\ \mathrm{mA}$), $V_{\mathrm{nl,max}}$ vs. $T$.  Both components become measurable only above $\sim 230\ \mathrm{K}$.  Fig. \ref{Fig5}c) indicates that this temperature correlates with the disappearance of spin freezing in the $a$-YIG.  Here we plot $\Delta m$ vs $T$, the component of magnetization due to the $a$-YIG film deposited on a Si-N coated Si substrate (isolation of this component from total measured SQUID magnetization is described in supplementary materials) for both zero-field-cooled (red symbols) and field-cooled (black symbols) states using a magnetic field of $5000\ \mathrm{Oe}$.  In contrast to existing literature on $a$-YIG, we see splitting of these curves at two temperatures, near the expected $50$ K peak in the ZFC curve, and at a temperature nearly equal to the observed onset of spin transport effects.  This suggests that spin transport occurs in the presence of disorder and strong spin correlations but only when sufficient thermal energy is available to overcome spin freezing.  


As seen in non-local spin transport in crystalline YIG, Figure \ref{Fig4} indicates a sharp drop in spin signal with increased separation between Pt strips, $L$, for both the linear and non-linear components of $V_{\mathrm{nl}}$.  These data do not fit a simple exponential dependence. We require more data to effectively probe existence of  diffusive and relaxation regimes\cite{CornelissenNatPhys2015}, and more detailed examination of separation dependence is ongoing.  We are able to clarify that any thermal component to $V_{\mathrm{nl}}$ here is small, and has a different dependence on $L$, further evidence that electrical effects dominate spin transport in the experiment on the substrate.


\begin{figure}
\includegraphics[width=5.8in]{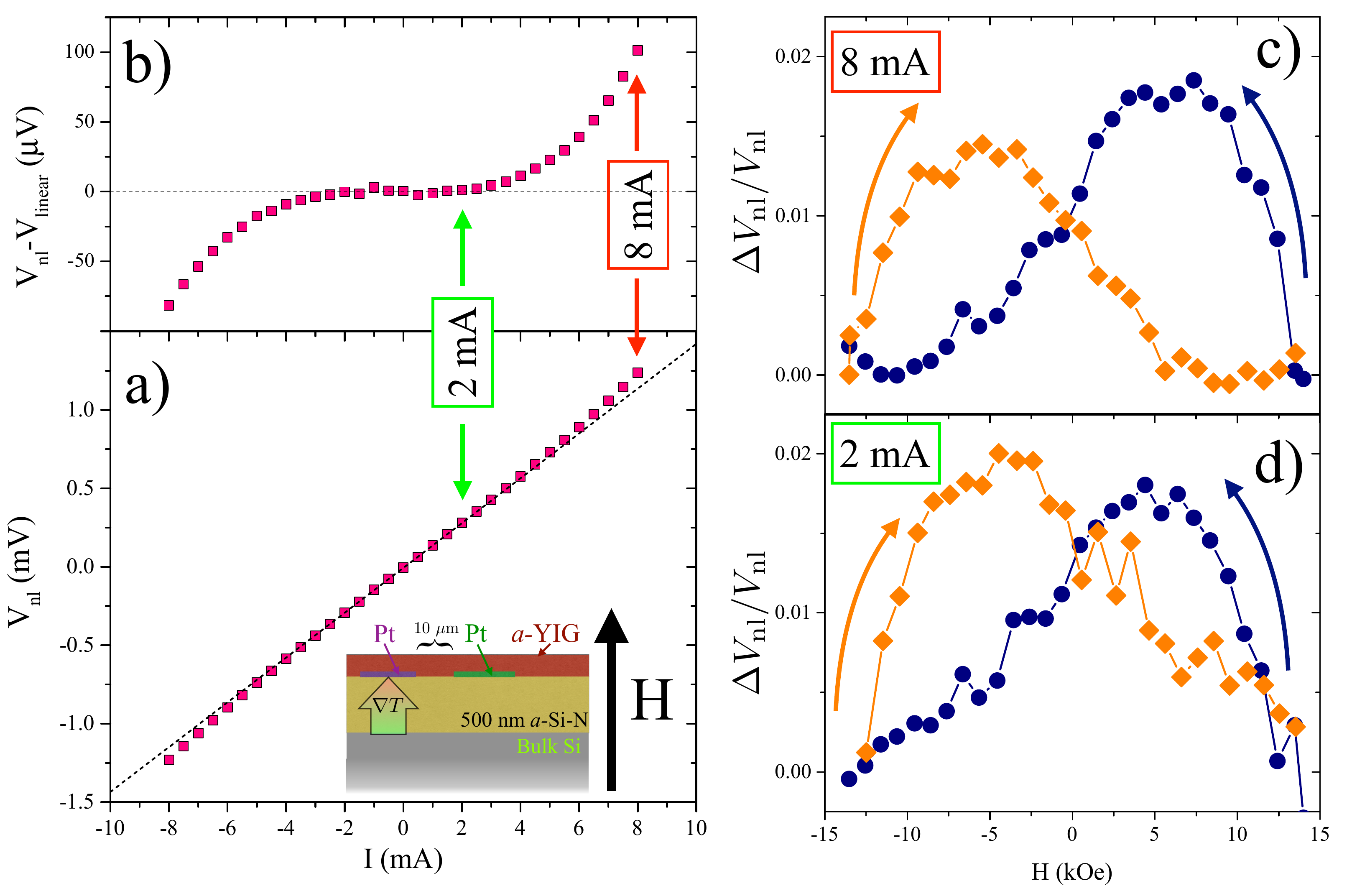}
\caption{Dependence of $V_{\mathrm{nl}}$ on applied field.  Here $H$ up to $14,000\ \mathrm{Oe}$ was applied perpendicular to the substrate as shown inset in \textbf{a)}, which shows $V_{\mathrm{nl}}$ vs. $I$, here measured in ambient conditions, and displaying the same linear and non-linear contributions as earlier Figs. \textbf{b)} The non-linear component isolated by subtraction of the linear term.  These clarify that when biased at $I=2\ \mathrm{mA}$ the signal is dominated by the linear term, where at $8\ \mathrm{mA}$ the non-linear term contributes.  Panels \textbf{c)} and \textbf{d)} show that at both bias points, clear field dependence is observable, with similar relative magnitude and trends. }
\label{Fig6}
\end{figure}

In disordered spin systems, even above any freezing temperature, strong AF spin correlations typically lead to small magnetic susceptibility and very large saturation fields.  This is the case for $a$-YIG, where $M$ is a very small fraction of either the saturation magnetization of crystalline YIG or of the even larger estimated $M$ of free Fe atoms at the same density.  Despite achieving a magnetization less than $10\%$ of the YIG value (described further in supplemental materials), as shown in Fig. \ref{Fig6} there is an observable effect of applied field on $V_{\mathrm{nl}}$.  Figure \ref{Fig6}a) shows $V_{nl}$ as a function of applied $I$ for the substrate-supported $a$-YIG film, here measured in air at room temperature.  Fig.\ \ref{Fig6}b) isolates the nonlinear component, which is near zero for $I\leq 2\ \mathrm{mA}$.  We applied fields up to $14$ kOe perpendicular to the film, large enough to have saturated $M$ and completely eliminated spin transport in crystalline YIG.\cite{GoennenweinAPL2015}  Figs.\ \ref{Fig6}c) and d) show $\Delta V_{\mathrm{nl}}/V_{\mathrm{nl}}$ vs. $H$ for $I=8\ \mathrm{mA}$ and  $I=2\ \mathrm{mA}$, respectively, and show that both the linear and non-linear regimes react to $H$ in a similar manner as expected if the field dependence arises from magnetic-field dependent properties of the medium.      Here $\Delta V_{\mathrm{nl}}/V_{\mathrm{nl}}=\left( V_{\mathrm{nl}}(H)-V_{\mathrm{nl}}(H=14\ \mathrm{kOe})\right)/V_{\mathrm{nl}}(H=14\ \mathrm{kOe})$.  Reduction from maximum $H$ does increase the signal, with the zero field values slightly reduced from a peak that occurs at intermediate fields.  The slight asymmetry in the peak value when starting from either value of maximum field is likely due to error on the subtraction procedure.  The small shifts in $V_{\mathrm{nl}}$ are consistent with the small shift in total magnetization achieved here.  Despite the small size, this field dependence is strong evidence that $V_{\mathrm{nl}}$ for $a$-YIG relies on spin transport.

The recent theories that explain spin transport in antiferromagnetic insulators\cite{KhymynPRB2016,BaltzarXiv2016,RezendePRB2016} invoke a well-defined antiferromagnetic magnon spectrum that is either absent or substantially modified in the case of a truly disordered system as we use here.   The magnon spectrum of disordered magnets has been rarely explored in the past, though existing work suggests an analogy to phonon spectra in glassy systems.\cite{HuberSSC1974}
Vibrational modes of amorphous systems certainly exist, and a long history of study shows that whether called a phonon or given a more specific name (such as propagon), heat transport via a broad spectrum of vibrational excitations is possible in amorphous systems \cite{WingertSST2016}.  Recent work\cite{BraunPRB2016,LarkinPRB2014,LiuPRL09,SultanPRB2013} shows that this transport is often surprisingly efficient, with long phonon mean free paths despite the disorder.  Our work is the first indication of similar effects for spin transport via magnetic correlations in a disordered system.  

In fact, use of a disordered system has potential advantages for magnonics.  Two traditional challenges for magnonic materials are the presence of a gap in the magnon spectrum and the highly anisotropic nature of the magnon transport introduced in a crystal \cite{KruglyakJPhysD2010}.  Neither should occur in a disordered system.  A central question is if spin transport effects in disordered systems persist over long enough length scales to be useful technologically.  The data shown here proves emphatically that they do.  The easy compatibility of the $a$-YIG material in any device process is also compelling, suggesting a potential paradigm shift in materials science for magnon transport. 


\section{Methods}

\subsection{Device Fabrication}

Thermal isolation platforms are fabricated from $500$ nm thick Si-N coated Si wafers (nominally $500$ microns thick) via bulk Si micromachining via an anisotropic Si etch in TMAH performed after defining the platform geometry by patterning the Si-N layer, which acts as a hard mask for the Si etch.  Before this etch, an evaporated Cr/Pt ($10$ nm/ $40$ nm) film is patterned via liftoff, which forms the leads, heaters, and thermometers used in thermal experiments.  Additional fabrication details are available elsewhere.\cite{SultanPRB2013}

\subsection{$a$-YIG Deposition}

 $200$ and $100$ nm thick $a$-YIG films were sputtered onto amorphous Si-N thermal isolation platforms and 1 cm $\times$ 1 cm blank Si-N coated silicon substrates from a stochiometric YIG target in argon gas.  The substrates were held near room temperature, and the material grown at $\sim 0.5$ nm/min.  Other parameters follow the sputtering step described in Ref.\ \onlinecite{ChangIEEEMagLett2014}.

\subsection{X-ray Diffraction}

X-ray diffraction data were collected in the Bragg-Brentano symmetrical $\theta - 2\theta$ reflection geometry by using Cu$_{\mathrm{K\alpha}}$ characteristic energy ($8$ keV), excited at $30$ kV and $30$ mA. Reflected intensity was scanned by a proportional detector every $0.05^{\circ}$ in the angle $2\theta$ for $10$ s per step.   XRD was measured on a $200$ nm thick $a$-YIG layer deposited on a $500$ nm thick Si-N coated Si substrate, where the $a$-YIG was grown in the same deposition as films on thermal isolation platforms tested for spin transport and magnetization.  For polycrystalline YIG the spectrum was normalized to the (420) peak, which had a raw value of  $13,000$ counts.   

\subsection{Non-local Transport Measurements}

After depositing this $a$-YIG film on our thermal isolation platform 
we carried out two series of non-local spin transport experiments.  We first drove a current $I$ down the length of a Pt lead traversing both legs of one Si-N island while measuring a voltage on a parallel but totally separate Pt strip with a gap of $\sim 9$ microns.  The total length of these strips is greater than $2$ mm and both are suspended on the Si-N membrane for this entire length.   We measured this non-local signal as a function of temperature from cryostat temperatures of $\sim 80-350$ K.  When large $I$ is driven through the Pt wire on the suspended membrane significant heating occurs, which we can monitor using the separate thermometer patterned on the island.  When heated, a thermal gradient in the suspended structure is generated that is nearly entirely confined to the plane of the film by the essentially 2d nature of the structure.  The second set of experiments explores the opposite regime of thermal gradient by performing the non-local injection and detection entirely supported by the substrate  (Figs.\ \ref{Fig5} and \ref{Fig4}).  There the overwhelming heat sink provided by the bulk substrate forces the thermal gradient to be perpendicular to the plane of the film.  

For all zero-field transport measurements the platforms are mounted in gold-coated OFHC copper sample mounts, leads are ultrasonically wire-bonded to custom circuitboards, and a radiation shield installed ensuring an isothermal sample environment.  This mount is attached to the cold finger of a sample-in-vacuum cryostat.  In all experiments not specifically stated to use exchange gas, vacuum of $10^{7}$ Torr or better is maintained around the sample.  Transport measurements here use standard computer-controlled source-meter equipment.  Voltage is measured as a function of applied current, and analyzed to determine its components.  Linear terms in $V_{\mathrm{nl}}$ are determined by fitting the low $I$ portion of the curve ($\leq 100\ \mathrm{GA/m}^{2}$), and in the case of substrate measurements, the thermally driven $\propto I^{2}$ term plotted in Fig. \ref{Fig4}b is estimated by taking the average value of the maximum and minimum $I$ data points, which show a slight but reproducible asymmetry indicating the presence of this parabolic contribution.  In the case of measurements on membranes, the temperature of the suspended islands are measured using entirely separate patterned thermometers connected to an AC lock-in based resistance bridge.  We also separately measured the (very high) direct resistance of the devices at various temperatures to ensure these are due to spin transport (discussed in more detail in Supplementary Materials).   

Field-dependent measurements were performed in ambient conditions, with the sample placed between the $~10\ \mathrm{cm}$ diameter pole pieces of an electromagnet with a gap of $<1\ \mathrm{cm}$.  Due to the strong temperature dependence of $V_{nl}$ as well as any thermoelectric background voltages, voltage measurements recorded while cycling a set bias current on and off were averaged over several cycles to reduce backgrounds.  The small field dependence is observable over the remaining background, which is linear over short enough time periods, and removed with a simple linear fit. 

\section{Acknowledgements}

We thank A. Hojem for helpful discussions and assistance in the lab, D. Schmidt for assistance with optical imaging, the NIST Boulder magnetics group for access to the SQUID magnetometer and advice, X. Fan and A. Humphries for deposition of the SiO$_{2}$ film, and J. Nogan and the IL staff at CINT for guidance and training in fabrication techniques.   D.W. and B.L.Z. gratefully acknowledge support from the NSF  (DMR-1410247). This work was performed, in part, at the Center for Integrated Nanotechnologies, an Office of Science User Facility operated for the U.S. Department of Energy (DOE) Office of Science by Los Alamos National Laboratory (Contract DE-AC52-06NA25396) and Sandia National Laboratories (Contract DE-AC04-94AL85000). The growth of the YIG films at CSU was supported by the SHINES, an Energy Frontier Research Center funded by the U.S. Department of Energy, Office of Science, Basic Energy Sciences under Award SC0012670. 

\section{Author Contributions}

Thermal isolation platforms were designed by D.W. and B.L.Z, and fabricated, measured, and analyzed by D.W. under supervision of B.L.Z.  $a$-YIG films were deposited by T.L. under supervision of M.W.  XRD on films and YIG substrates was performed and analyzed by D.B. FEM thermal calculations were performed by D.W. with consultation and input from B.L.Z.  B.L.Z. initiated the study with consultation from M.W.  D.W. and B.L.Z. wrote the manuscript with contributions from all authors.

\clearpage
%


\clearpage
\setcounter{equation}{0}
\setcounter{figure}{0}
\setcounter{table}{0}
\setcounter{page}{1}
\makeatletter
\renewcommand{\theequation}{S\arabic{equation}}
\renewcommand{\thefigure}{S\arabic{figure}}
\renewcommand{\bibnumfmt}[1]{[S#1]}
\renewcommand{\citenumfont}[1]{S#1}

\textbf{\Large Supplemental Information for ``Long-distance spin transport in a disordered magnetic insulator"}

\section{Additional Background Measurements: SiO$_{2}$, CNT}

\begin{figure}
\includegraphics[width=3.38in]{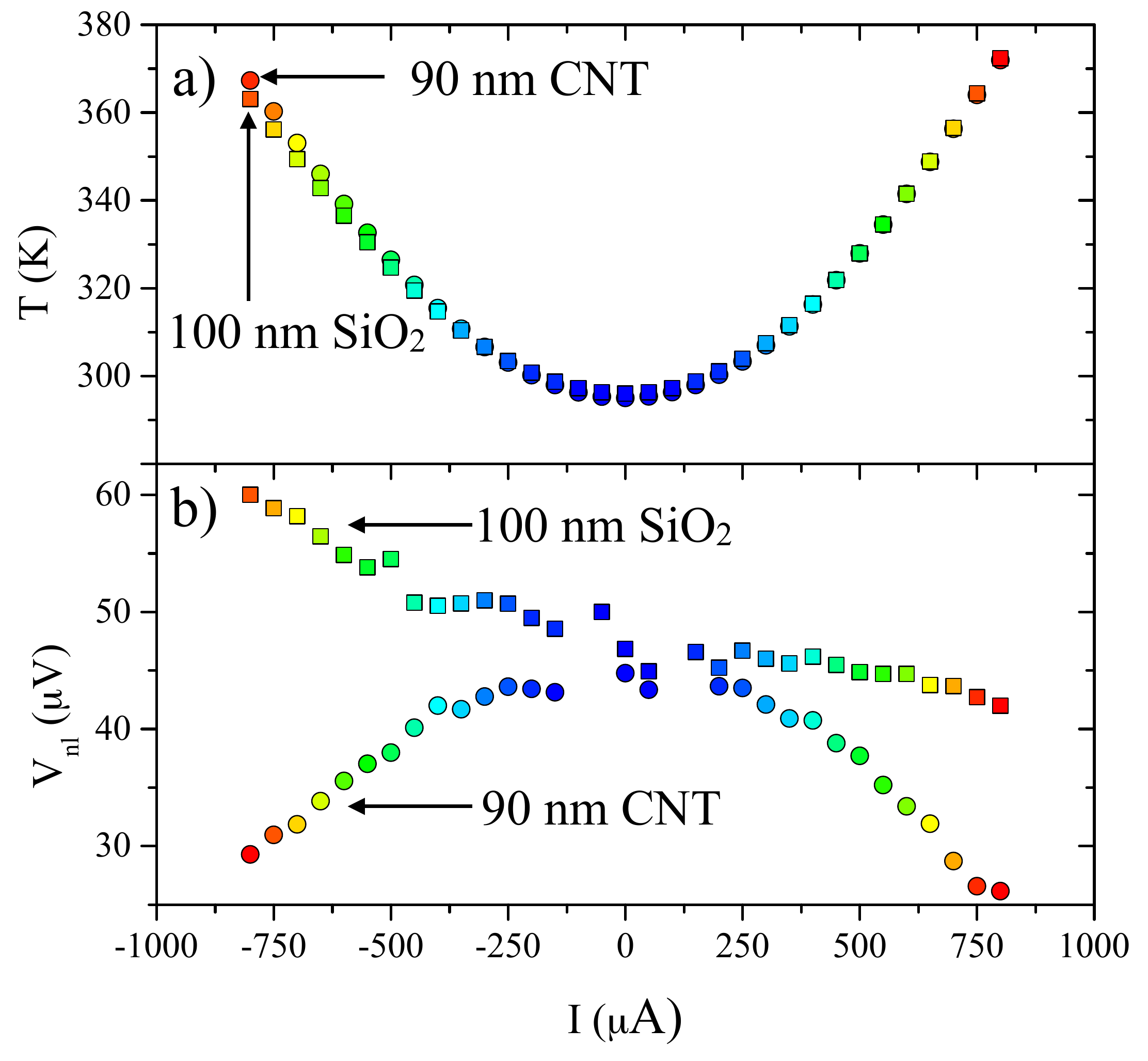}
\caption{\textbf{a)} Temperature and \textbf{b)} Non-local voltage, $V_{\mathrm{nl}}$, vs. current applied to the Pt injector strip as described in the main text.  Boxes show results for a thermal isolation platform coated with a $100$ nm-thick film of sputtered SiO$_{2}$, and circles for a platform coated with a $90$ nm-thick doped carbon nanotube network thin film (described in more detail elsewhere\cite{AveryNatEn2016}).  The fill color indicates the temperature in both panels, and the same color scale is used for all data sets.}
\label{SiO2CNT}
\end{figure}

In addition to the ``blank" Si-N membrane measurement shown in Fig.\ \ref{aYIGdata}c), we verified the absence of a spin signal in two other films, an insulating (but non-magnetic) $100$ nm sputtered SiO$_{2}$ film and a $90$ nm doped carbon nanotube (CNT) network film that we previously characterized in detail for thermal, electrical, and thermoelectric properties.\cite{AveryNatEn2016}  Results from non-local transport experiments on these samples are shown in Fig.\ \ref{SiO2CNT}.  As in Fig.\ \ref{aYIGdata} the top panel shows temperature measured by the separate island thermometer, while the lower panel shows the measured non-local voltage, $V_{\mathrm{nl}}$, both as a function of applied current, $I$, to the injector strip.  The resulting heating (free from non-monotonic behavior associated with spin excitations, as was the bare Si-N platform) is nearly identical between the two platforms.   

\begin{figure}
\includegraphics[width=3.38in]{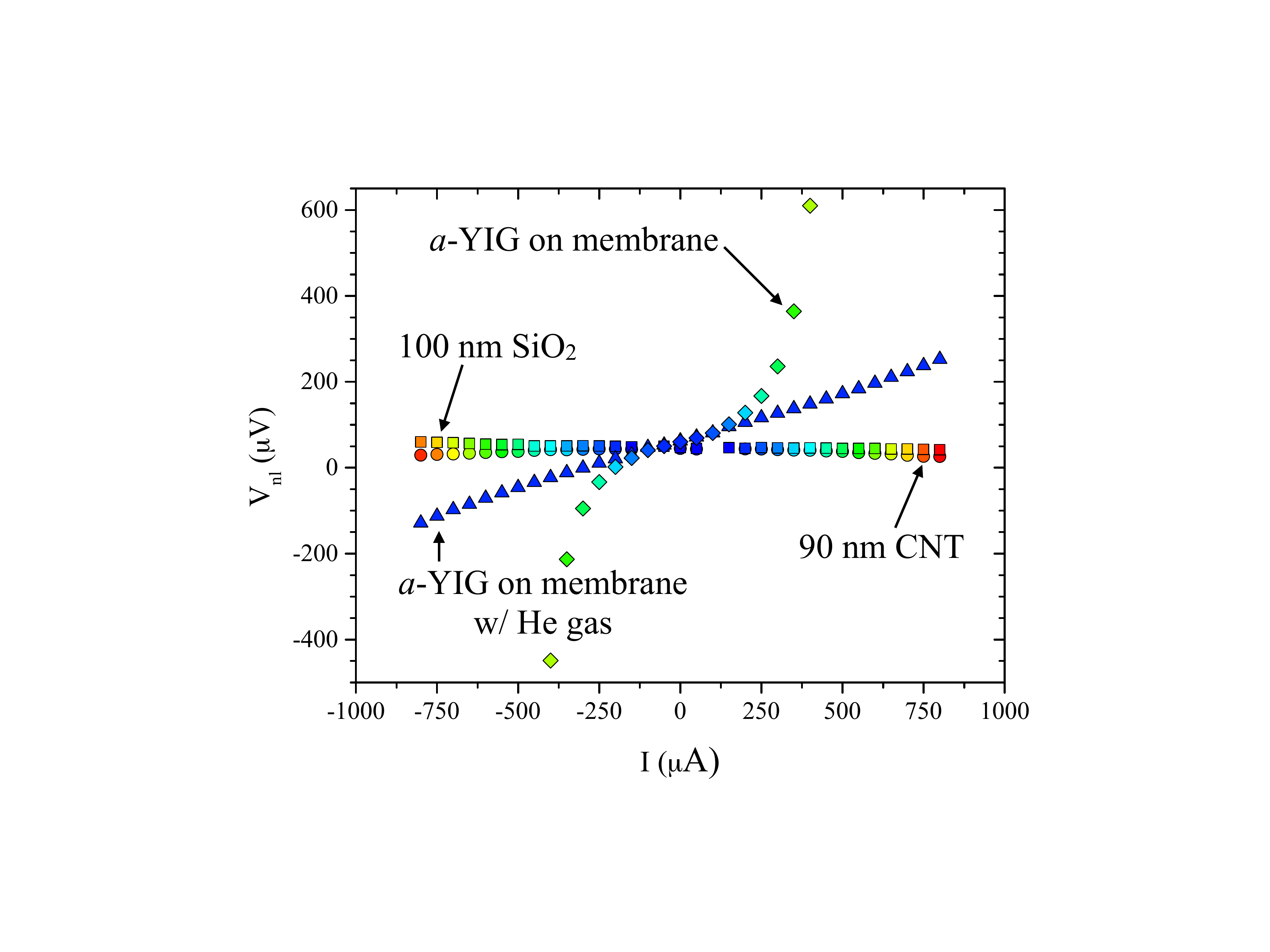}
\caption{The same two data sets shown in Fig.\ \ref{SiO2CNT} compared on a larger scale to the $a$-YIG non-local data for the membrane with He exchange gas, and in vacuum.  This clarifies the small scale of the background voltages in Fig.\ \ref{SiO2CNT}, even compared to the smaller, largely linear spin signal generated when the in-plane thermal gradient is cancelled.  This plot also clarifies the small $V_{\mathrm{nl}}$ term $\propto I$ in the $a$-YIG with an in-plane thermal gradient, which is not visible on the full scale plots shown in Fig.\ \ref{aYIGdata}. }
\label{SiO2CNTmem}
\end{figure}

In both platforms, small non-local voltages are measurable, though orders of magnitude smaller than the signals seen in $a$-YIG and with very different dependence on $I$.  $V_{\mathrm{nl}}$ for the SiO$_{2}$ film shows a roughly linear behavior, possibly due to a drift of the base temperature for this measurement.  
$V_{\mathrm{nl}}$ for the CNT film is predominantly proportional to $I^{2}$ with a constant offset voltage, obvious from the $I=0$ value.  Both these terms have a simple thermoelectric origin, with the constant offset due to the small temperature difference between the sample region and the room temperature electronics, and the $\propto I^{2}$ term likely due to a thermal gradient that develops in the platform as current is driven down the Pt injector strip.  We previously measured the Seebeck coefficient of this CNT film to be $S\simeq 70\ \mathrm{\mu V/K}$ in this temperature range,\cite{AveryNatEn2016} suggesting that a total temperature difference across the sample region of approximately $0.2\ \mathrm{K}$ develops at the maximum applied $I$.  This is $\simeq 0.3\ \%$ of the maximum temperature of the island,  likely due to small asymmetries that exist in the thermal platform or the sample itself, and is in line with background signals seen in measurements of other transverse thermovoltages.\cite{AveryPRL2012,MasonThesis}  Note that this CNT film also has an easily measurable \emph{electrical} conductivity, such that the charge resistance between the two Pt strips is $\sim100-200\ \mathrm{\Omega}$. 
This much more efficient charge transport channel than exists in the $a$-YIG does NOT lead to the sign-reversing signal we identify with spin transport.  
To clarify this, we plot $V_{\mathrm{nl}}$ for these two backgrounds with the data for suspended $a$-YIG both in vacuum and with exchange gas in Fig.\ \ref{SiO2CNTmem}.  Here even the comparably smaller linear term that dominates in $a$-YIG nonlocal transport when the in-plane thermal gradient is cancelled is much larger than either background measurement.  When a significant in-plane thermal gradient is applied, the non-linear part of $V_{\mathrm{nl}}$ dominates and is many orders of magnitude larger (Fig.\ \ref{SiO2CNTmem} shows only a small portion of this data). 

\section{Consideration of leakage current}

\begin{figure}
\includegraphics[width=5.0in]{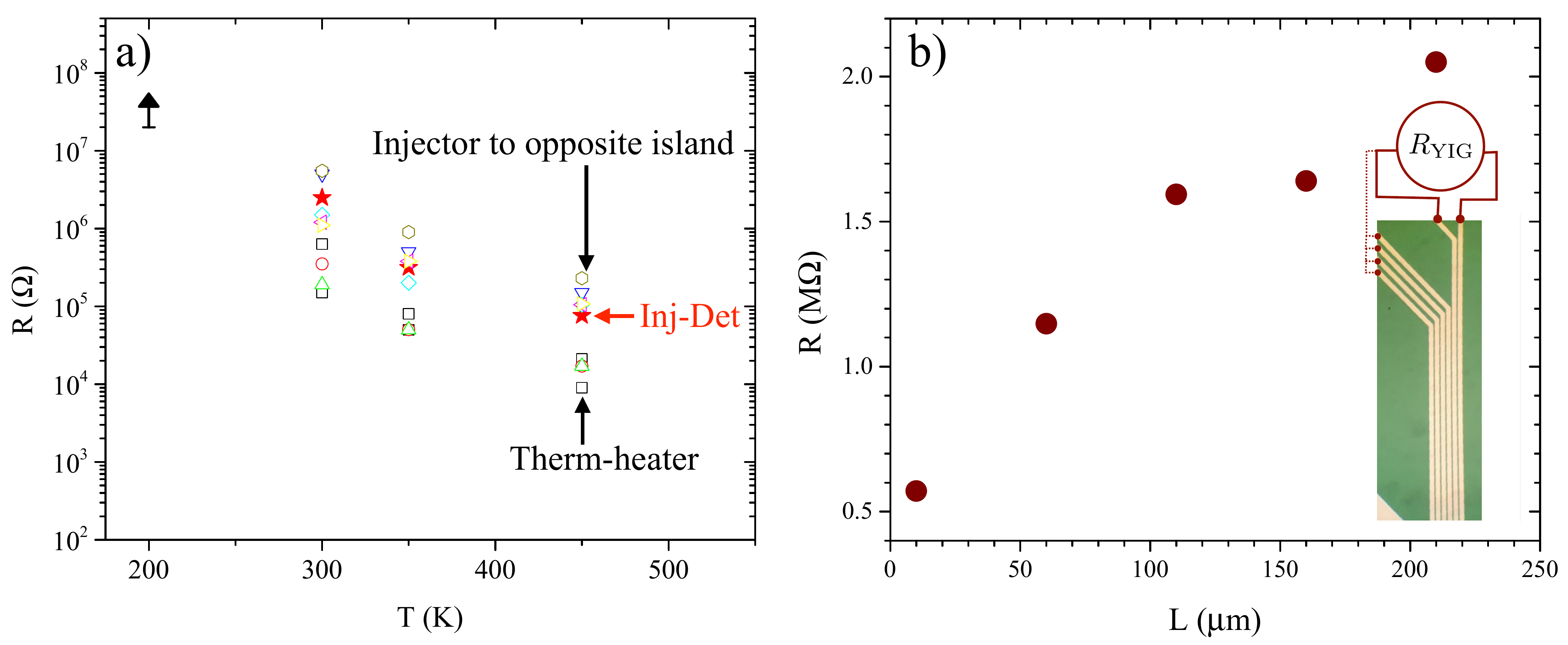}
\caption{\textbf{a)}Two-point resistance measurements through the 200 nm thick $a$-YIG layer deposited on the thermal isolation platform between various metal features all show very large values of resistance, and all show the same roughly exponential behavior as a function of $T$.  At $200$ K and below $R$ became too large for the standard digital voltmeter used here to measure. These results are all consistent with charge flow through a large band gap semiconducting layer, taking the geometry into consideration. \textbf{b)} Two-point resistance as a function of separation for the strips used in Fig.\ \ref{Fig4} shows the simple, roughly linear increase with separation expected from the geometry. }
\label{Rleak}
\end{figure}

Though the CNT data is strong evidence that a conducting path between Pt strips does not lead to the signals we associate with spin transport, we also checked carefully for any significant charge conductance through the $a$-YIG film. Figure \ref{Rleak} details 2-point resistance measurements between metal features on the thermal isolation platform for various temperature.  All of these values are orders of magnitude larger than the ``leakage" path present in the CNT film discussed above.  To compare to measured non-local voltages generated by spin transport, one can calculate $R_{\mathrm{spin}}=V_{\mathrm{nl}}/I$ for the linear and non-linear components to $V_{\mathrm{nl}}$.  These values fall in the range of $10-100\ \mathrm{\Omega}$, many orders of magnitude less than measured charge resistance. 

\begin{figure}
\includegraphics[width=\linewidth]{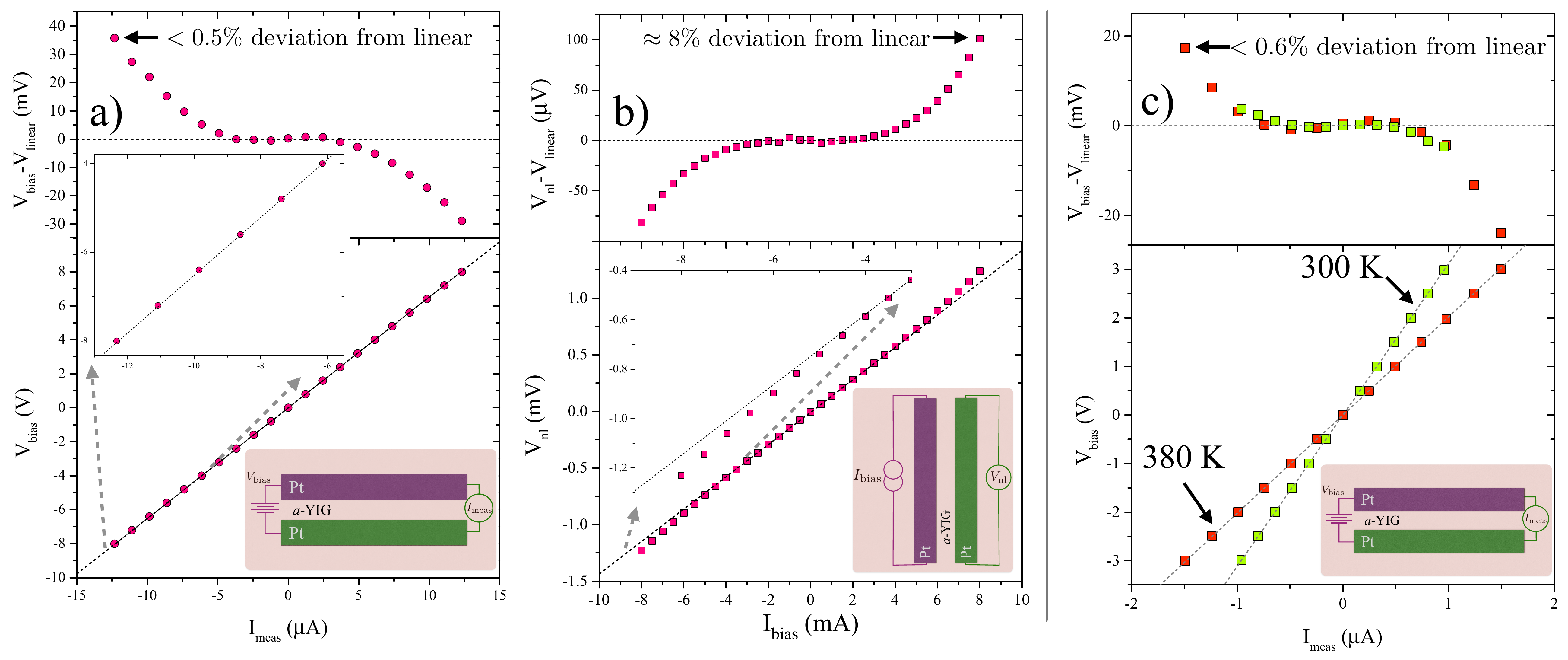}
\caption{Comparison of \textbf{a)} voltage-biased, measured current and \textbf{b)} standard current-biased non-local voltage (right) measurements.  Both \textbf{a)} and \textbf{b)} measurements are on the same Pt strips and are on the substrate.  Dashed lines in both main plots indicate a linear fit to the data near zero current, extrapolated across the range of data.  The top plots show the deviation from this linear curve by subtracting it from $V$.  A very small non-linearity under large voltage bias changes the $\approx 650\ \mathrm{k\Omega}$ effective resistance between the Pt strips by $<1\%$.  This small non-linear effect, likely the result of Schottky barriers formed at the Pt/$a$-YIG interface, cannot explain any of the non-local signals. \textbf{c)} The voltage-biased measurement for the the Si-N membrane shows similarly tiny non-linearity compared to the very large effects under current bias.}
\label{VbiasSub}
\end{figure}

To further eliminate the possibility of leakage currents or Schottky barrier effects contributing to non-local voltages we measure here, we have directly measured current flowing between the Pt strips under large bias voltages.  As the application of large currents to the Pt injector, with $R\sim 1000\ \mathrm{\Omega}$ in the case of the substrate-supported experiment, causes a large voltage drop across the entire length of the injector wire, some portion of the $a$-YIG between the Pt strips experiences an effective bias voltage up to potentially $8$ V.  Here we perform a more rigorous check, by applying $V_{\mathrm{bias}}$ up to 8 V between the strips across their entire length.  In Fig.\ \ref{VbiasSub}a we plot measured current, $I_{\mathrm{meas}}$ against $V_{\mathrm{bias}}$ (with measured current on the $x$-axis) to simplify comparison to the non-local IV curves for the substrate-supported Pt/$a$-YIG.  We do see a very small non-linearity under voltage bias, likely indicating the presence of Schottky barriers at the Pt/$a$-YIG interface.  However, as the overall effect here is to shift the effective resistance (the slope of this curve) by $<1\%$ at large voltage biases, this cannot explain the much larger non-linear voltage components seen when large bias current drives spin transport in the $a$-YIG.  

We note that though simplistic network models or finite-element calculations do allow that linear voltage components on the order of mV or less are possible in the non-local measurement due to leakage of charge through the $a$-YIG, there are two pieces of strong evidence against a charge-leakage origin for the linear term.  The first is that the linear component drops off dramatically with distance, while as shown in Fig.\ \ref{Rleak}b, the measured resistance through the $a$-YIG due to charge effects goes as expected from a simple increased length of the current path approximately linearly with distance.  Specifically, the slope of the IV curve drops by a factor of more than 20 between the $10\ \mu\mathrm{m}$ and $110\ \mu\mathrm{m}$ separations, while the $a$-YIG resistance increases by less than a factor of 3.  The second piece of evidence that argues against a charge leakage explanation, as described in more detail in the main text is the field-dependence of both the linear and non-linear terms.   

We have also tested the case of voltage bias as a function of temperature for the experiments on $a$-Si-N membranes.  Fig. \ref{VbiasSub}c) shows two voltage-bias experiments for two different membrane temperatures (where heat is added to the island heater for the $380$ K data).  Here current-biased experiments are not shown, but result in the very large non-linear contributions apparent in Fig.\ \ref{aYIGdata}.  Again under voltage bias we see very nearly linear responses, with non-linearly much less than $1\%$.  The leakage resistance is temperature dependent, but no large non-linear terms appear even at elevated temperatures.  We can furthermore again rule out any contribution of voltage leakage as a complete explanation of the linear term in non-local experiments with large current biases.  Consider that the leakage resistance through $a$-YIG drops by $\sim 60\%$ between $300$ and $380$ K, where the linear term in non-local measurements with large current bias increases by more than a factor of ten.  The non-linear voltage under current bias at a membrane temperature of $380\ \mathrm{K}$ is $350\times$ larger than the linear term, while the charge leakage under voltage bias changes by $<1\%$.   This clarifies that neither the linear nor non-linear components of the non local voltages arises from charge transport through the $a$-YIG. 
\section{SQUID Magnetometry of $a$-YIG films}

\begin{figure}
\includegraphics[width=3.38in]{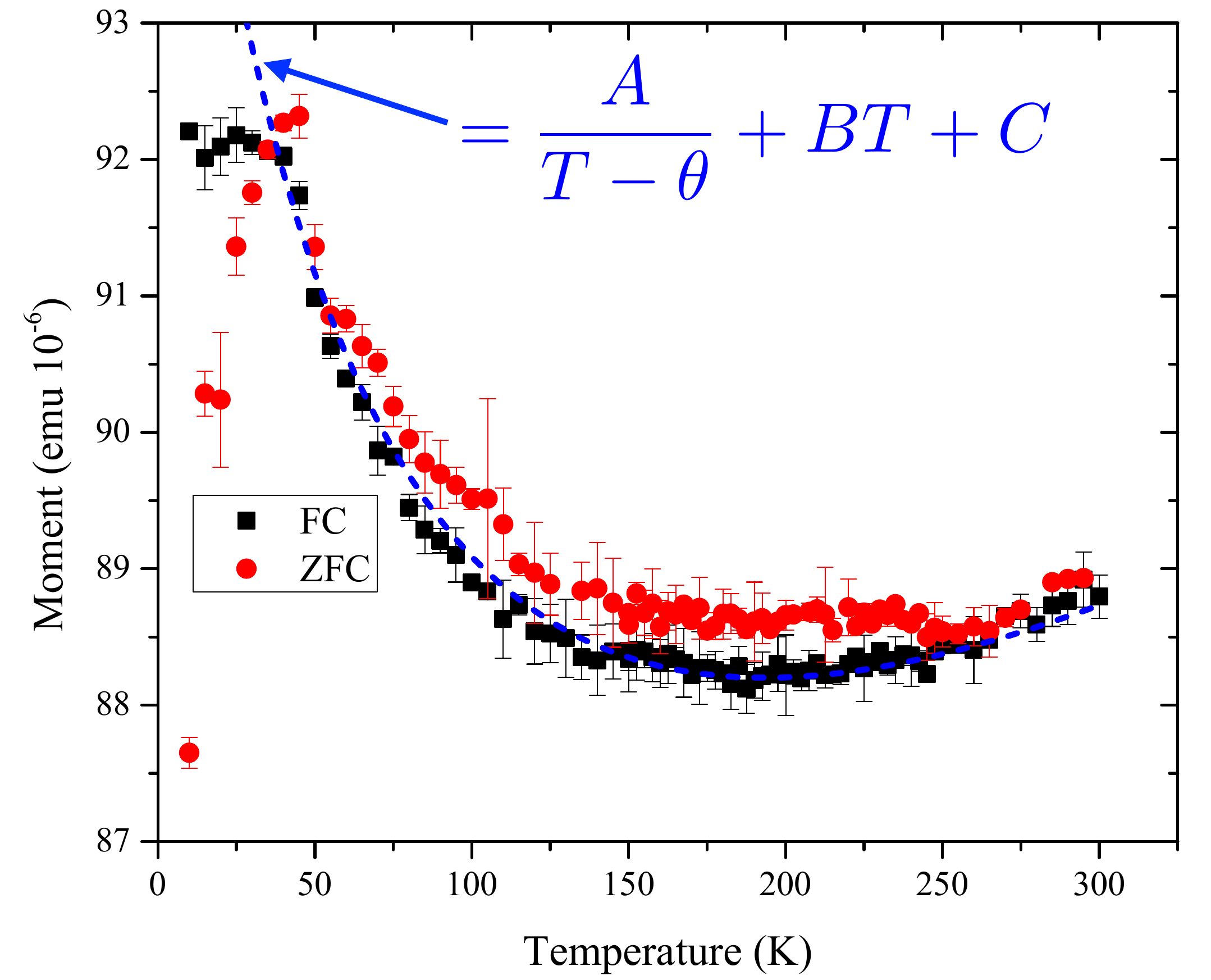}
\caption{Moment vs. $T$ in 5000 Oe applied field for both FC (black box) and ZFC (red circle) states.  The blue line includes three terms, with two background contributions as described in associated supplemental text.  Here $B=1.73\times10^{-8}\ \mathrm{emu/K}$ and $C=80.3\ \mathrm{\mu emu}$, these two terms are subtracted from both FC and ZFC data for the plot in the main text.}
\label{RAWsquid}
\end{figure}

We measured magnetization as a function of temperature of $a$-YIG using a Quantum Design MPMS SQUID Magnetometer.  As stated in the main text, 200 nm thick $a$-YIG was simultaneously deposited on Si-N thermal isolation structures and $1$ cm $\times$ $1$ cm Si-N coated Si substrates.  One of these substrates was cut into smaller pieces, placed in a gelatin capsule, held in place using cotton batting, and mounted in a drinking straw that was mounted in the magnetometer.  $M$ vs $T$ for first zero-field-cooled and then field-cooled conditions was measured from $5-300$ K in $5000$ Oe, an applied field much smaller than the typical exchange energy of the AF coupling in $a$-YIG.  The raw voltage was converted to moment using typical fitting procedures, and the resulting raw moment is shown in Fig.\ \ref{RAWsquid}.  Despite presence of both a significant temperature-independent paramagnetic background and a smaller, linear in $T$ paramagnetic background, all essential features of the $a$-YIG magnetization are obvious.  The large $T$-independent offset is most likely due to the background from the cotton batting,\cite{GarciaJAP2009} though more quantitative background determination is required to rule out an origin in iron clusters in the $a$-YIG itself that are too small to be observed in XRD as nanocrystallites.   The linear background term we associate with the temperature-dependent Van Vleck paramagnetism of the semiconducting Si substrate.\cite{NeySST2011,SawickiSST2011}  The plots in the main text subtract these two background terms, which gives the expected $M\propto1/(T-\theta)$ dependence above the $50$ K freezing temperature for the FC curve with $\theta$ on order of $-100\ \mathrm{K}$.   

\begin{figure}
\includegraphics[width=3.38in]{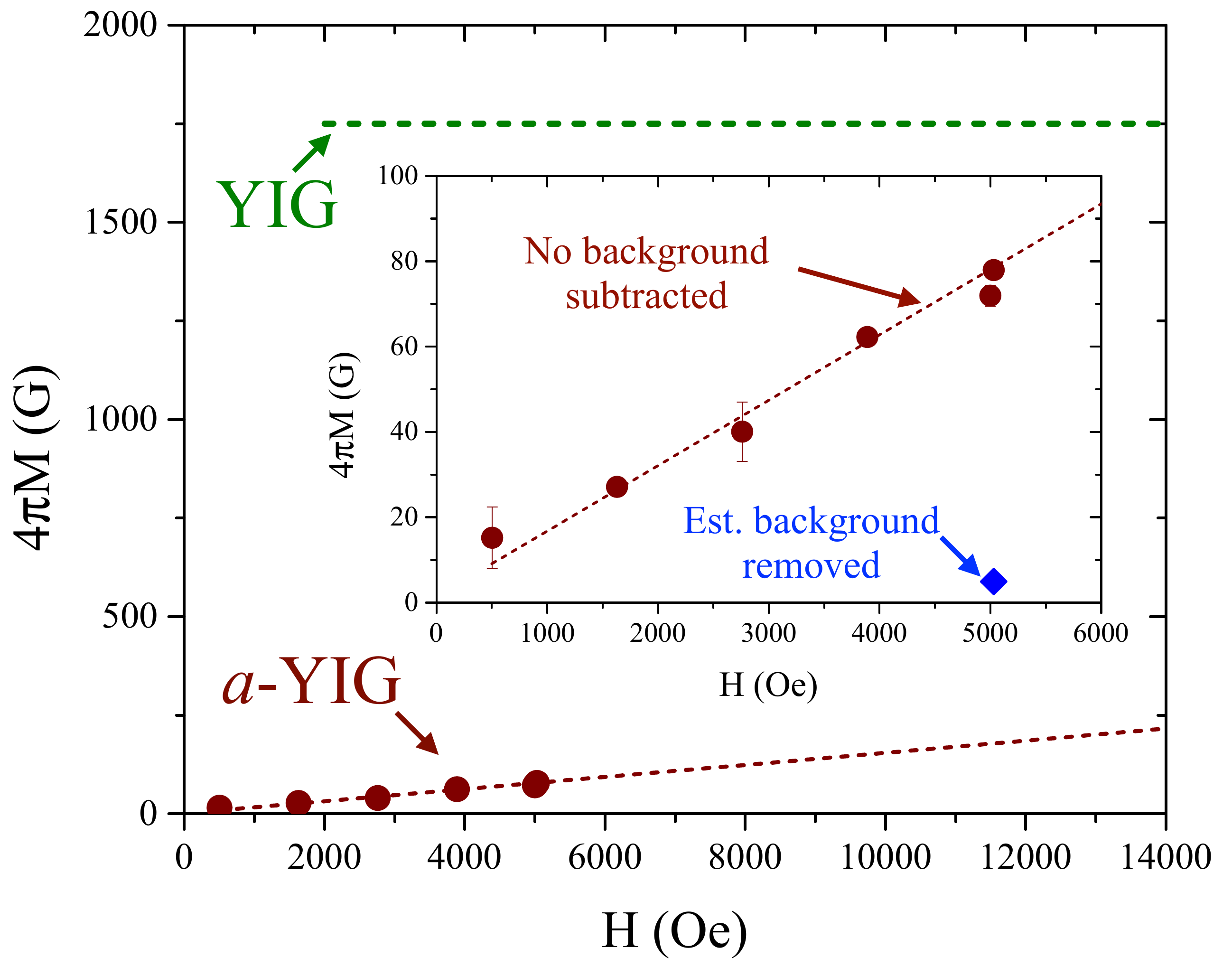}
\caption{Estimated $4\pi M$ vs. $H$ at $300$ K. Even taking the entire measured moment as the contribution of $a$-YIG (which is very unlikely but provides an upper limit) results in overall magnetization a small fraction of that commonly seen in ordered YIG.}
\label{MvsH}
\end{figure}

Figure \ref{MvsH} shows estimated magnetization vs. $H$ for the $a$-YIG film.  The main plot compares the upper bound of $4 \pi M$ for the $a$-YIG to the expected value for high-quality bulk and thin films of ordered YIG.  This upper bound is given by converting the entire measured moment of the sample, substrate, and mounting materials using the volume of the film.  This almost certainly very significantly overestimates the true magnetization of the $a$-YIG film itself.  Despite this, $4\pi M$ determined this way is at least an order of magnitude below the YIG value up to at least $5000$ Oe (well past the expected in-plane saturation), and remains linear with $H$ with large susceptibility, indicating no approach to saturation, which is a common feature of disordered magnetic systems even far above the freezing temperature.\cite{mydosh,Ternery,GdSiMag}  In such systems, very strong AF interactions are common, and here we expect the field required to approach saturation to be well out of the scale of typical laboratory superconducting magnets.

\section{Finite Element Thermal Modeling}

\begin{figure}
\includegraphics[width=\linewidth]{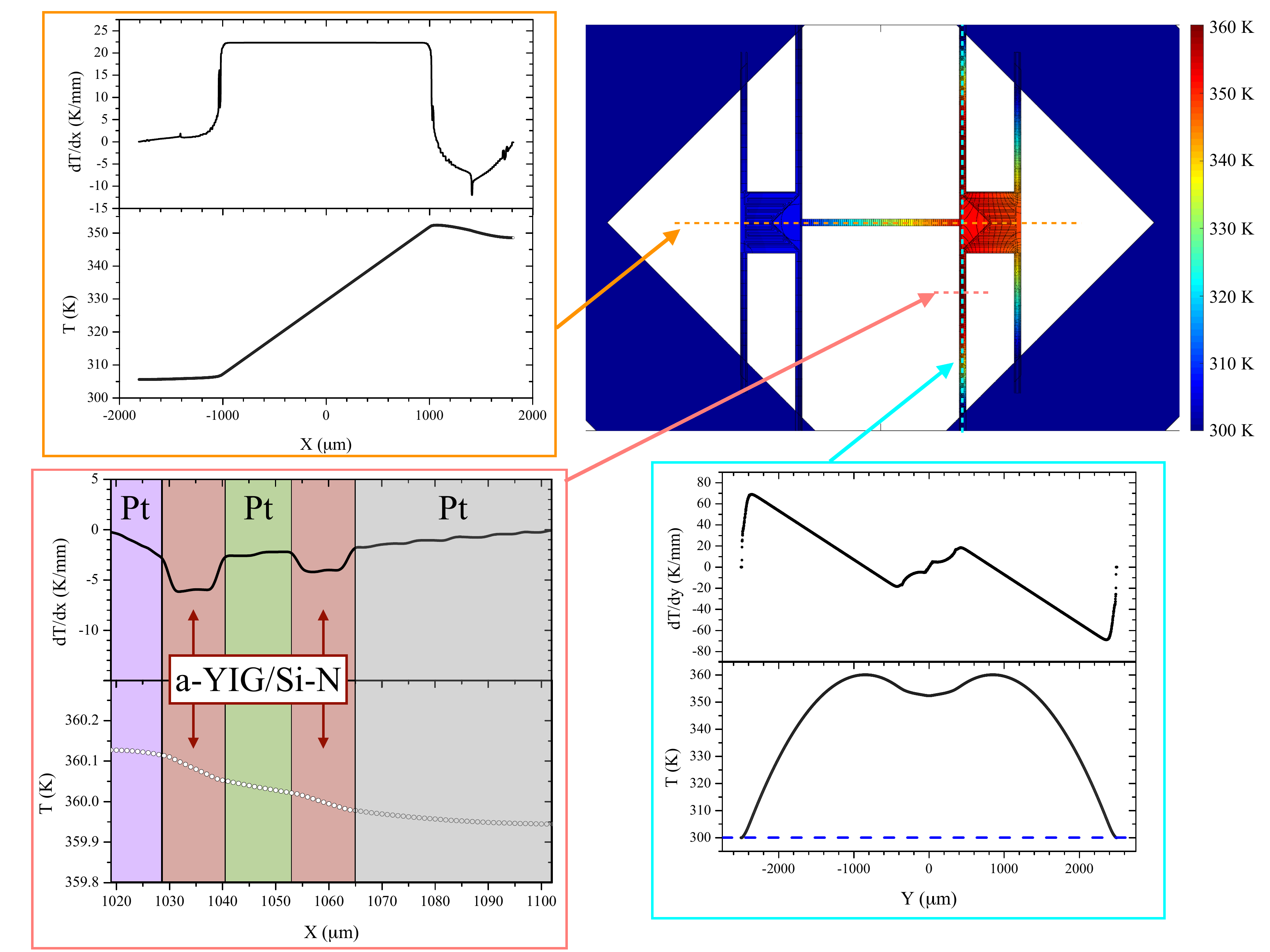}
\caption{Calculated thermal profile in the suspended platform.  \textbf{Upper Right:} Plot of $T(x,y)$ resulting from the 2d FEM simulation.  Heating is in response to the large current density passed through the Pt injector strip as described in the main text.  Dashed, colored lines indicate the locations of the 1d plots.  \textbf{Lower Right:} $T$ vs. $y$ and $dT/dy$ vs. $y$ along the leg between the Pt injector and detector strips.  \textbf{Upper Left:} $T$ vs. $x$ and $dT/dx$ vs. $x$ along the center of the bridge connecting the two islands. \textbf{Lower Left:} Zoomed view of the $T$ vs. $x$ and $dT/dx$ at the point of maximum $T$ along the leg.  Shaded boxes indicate the locations of Pt strips and Si-N/$a$-YIG areas, where the thermal gradient is largest as expected. }
\label{MembraneSupplement}
\end{figure}

We performed finite element modeling in 2d using a common commercially available software package \cite{MatLab}.  This package allows solution of the 2d heat flow equation (for our purposes limited to steady-state):
\begin{equation}
\frac{\partial}{\partial x} \left( k_{\mathrm{2D}}\left(x,y\right)\frac{\partial T\left(x,y\right)}{\partial x}\right)+\frac{\partial}{\partial y} \left( k_{\mathrm{2D}}\left(x,y\right)\frac{\partial T\left(x,y\right)}{\partial y}\right)=P_{\mathrm{2D}}\left(x,y\right),
\end{equation}
where $k_{\mathrm{2D}}=k \cdot t$ with $k$ the thermal conductivity (in $\mathrm{W}/\mathrm{m} \mathrm{K}$) of the constituent materials. In the case of models of the essentially 2d suspended structures, $t$ is taken to be the known thickness of each film.  Where two films overlap, $k_{\mathrm{2D}}$ is the sum of both contributions.  We also estimate out-of-plane gradients for the sample-on-substrate case by taking $t$ to be a uniform thickness (here $1\ \mathrm{\mu m}$) of the hypothetical cross-section.  As long as the heat flow is dominated by the bulk substrate so that in-plane thermal transport is negligible on long length scales, such a model gives a reasonable estimate of the out-of-plane thermal gradient at the Pt/$a$-YIG interface.  To match our experimental conditions for the cross-sectional simulations (sample in vacuum, with substrate clamped at the bottom to a thermal bath), we choose the Dirichlet boundary condition at the base of the Si substrate (fixing $T=300$ K), and Neumann boundary conditions elsewhere with no radiative or convective heat flow.   The 2d models of the membrane assume the Si frame is clamped to the base temperature, and Dirichlet boundary conditions are used at all edges of the Si-N structure. 

Values of the thermal conductivity of the Pt lead are estimated from the Wiedemann-Franz law, and $a$-YIG by using (low) values taken from the thermal platform measurements.  For the Si-N underlayer, which is critical for realistic modeling, we take the value $\sim 3\ \mathrm{W}/\mathrm{m}\ \mathrm{K}$ that we measure frequently for this Si-N using the suspended Si-N platforms\cite{SultanPRB2013}, and use literature values for Si thermal conductivity ($\sim2000\ \mathrm{W}/\mathrm{m}\ \mathrm{K}$) \cite{GlassbrennerPR1964}.  For simplicity we use temperature-independent thermal conductivity (which is most likely a good assumption for the sample-on-substrate models, but could introduce inaccuracy for the thermal platform measurements where large temperature differences occur), and also make the simplifying assumption that all Joule heat is dissipated evenly in the injecting Pt wire.    

For the membrane measurements we set $P_{\mathrm{2d}}$ dissipated in the injector strip by matching the temperature to that measured by the island thermometer.  For substrate measurements, the known current applied is converted to the appropriate volumetric power dissipation.  The FEM problem is then solved using an adaptive mesh with $> 5000$ nodes. The resulting solution for $T \left(x,y\right)$ for the membrane is shown in Fig.\ \ref{aYIGdata}f)  and in greater detail in Fig.\ \ref{MembraneSupplement}. Values from this solution are exported and a numerical derivative of this curve as a function of the appropriate dimension gives the thermal gradient.  

\section{Estimation of spin thermal conductance in $a$-YIG}

We can estimate the size of this new heat pathway by comparison to the bare Si-N thermal isolation platform.  There the effective thermal conductance of the legs connecting the central Si-N island to the thermal bath is typically $K_{\mathrm{L,eff}}=P/\Delta T\sim2.4\ \mathrm{\mu W/K}$ near $300\ \mathrm{K}$, where $P$ is the power dissipated and $\Delta T$ the resulting temperature difference across the leg.  Achieving $\Delta T\simeq150\ \mathrm{K}$ as shown for the bare Si-N platform in Fig.\ \ref{aYIGdata}f) then requires an average power dissipation of $P\simeq360\ \mathrm{\mu W}$.  The reasonable assumption that the same applied $I$ in the $a$-YIG coated platform causes the same average applied $P$ allows the estimation for the much smaller temperature difference caused by the addition of the spin excitation thermal conductance channel once dynamics are excited in $a$-YIG such that $K_{\mathrm{L,eff}}=(360\ \mathrm{\mu W})/(70\ \mathrm{K})\simeq5\ \mathrm{\mu W/K}$, suggesting the spin excitations contribute a roughly equal heat conduction to the existing Si-N leg with its Pt leads.  Simple estimates based on the $a$-YIG film geometry indicate a spin thermal conductance $k_{\mathrm{spin}}>100\ \mathrm{W/m\ K}$, orders of magnitude larger than the $\sim 1\ \mathrm{W/m\ K}$ \emph{total} thermal conductivity of the $a$-YIG film seen in our measurements with no applied charge current in the Pt strip, and on the order of electronic thermal conductivities seen in polycrystalline metal films.\cite{AveryPRB2015}

%

\end{document}